\documentclass[12pt]{iopart}

\usepackage{iopams,cite} 

\usepackage{graphicx} 
\begin{document}

\title[Anomalous transport of subdiffusing cargos by single kinesin motors ]{Anomalous transport of 
subdiffusing cargos by single kinesin motors:
the role of mechanochemical coupling and anharmonicity of tether}

\author{Igor Goychuk}

\address{Institute for Physics
and Astronomy, University of Potsdam, Karl-Liebknecht-Str. 24/25, 14476
Potsdam-Golm, Germany}
\ead{igoychuk@uni-potsdam.de}
\vspace{10pt}
\begin{indented}
\item[] 
\end{indented}

\begin{abstract}
Here we generalize our previous model of molecular motors trafficking subdiffusing 
cargos in viscoelastic cytosol by (i) including mechanochemical coupling between  
cyclic conformational fluctuations of the motor protein driven by the reaction of ATP hydrolysis 
and its translational motion within the simplest two-state model of  hand-over-hand motion
of kinesin,  and also (ii) by taking into account the anharmonicity of the tether between the motor 
and cargo (its maximally possible extension length). It is shown that the major earlier 
results such as occurrence of normal versus anomalous transport depending
on the amplitude of binding potential, cargo size and the motor turnover frequency
not only survive in this more realistic model, but the results also look very similar
for the correspondingly adjusted parameters.
However, this more realistic model displays a substantially 
larger thermodynamic efficiency due to a bidirectional mechanochemical coupling.
For realistic parameters, the maximal thermodynamic
efficiency can be transiently about 50\% as observed for kinesins, and even larger, 
surprisingly also in
a novel strongly anomalous (sub)transport regime, where the motor enzymatic turnovers
become also anomalously slow and cannot be characterized by a turnover rate.
Here anomalously slow dynamics of the cargo enforces anomalously slow cyclic 
kinetics of the motor protein. 

\end{abstract}

%
%
%
%
%

\section{Introduction}

The problem of how molecular motors can operate and realize transport 
in such a crowded environment  as  cytosol of biological
cells \cite{Pollard,McGuffee,Luby} came only recently in the limelight of 
attention \cite{Caspi,Goychuk10,Robert,BrunoPRE,GKh12a,KhG12,
PLoSONE14,PCCP14,Bouzat}. 
Indeed, numerous recent experiments
reveal that submicron particles like various endosomes and organelles, mRNA molecules, ionic channels 
and even smaller nanoparticles diffuse passively anomalously slow with mean-square distance
growing sublinearly in time rather than simply diffuse (linear growth) on the relevant mesoscopic
time and spatial scales 
\cite{Caspi,Robert,Guigas,Saxton,Qian,Yamada,Seisenberg,Tolic,Golding,Szymanski,Bruno,Jeon11,
Jeon,Barkai,Tabei,Hofling,Taylor,Weber,WeissM,Harrison}. 
Molecular motors such as various kinesins are thus 
indispensable for delivering such and similar 
cargos e.g. along axons of neuronal cells \cite{Hirokawa}.
In a two-state flashing ratchet model of molecular motors with position-independent switching 
rates it has been shown that a power stroke like operation mechanism can perfectly overcome
subdiffusion slowness and result into a highly efficient normal transport characterized by mean transport
velocity \cite{PLoSONE14,PCCP14}. However, the very possibility to realize such a normal 
active transport of (passively)
subdiffusing cargos presents a highly nontrivial issue. It depends on (i) the binding strength of 
motor protein to the microtubule providing a transport highway, (ii) cargo size, (iii)
motor operating frequency, (iv) external force directed again the processive
motion of the motor \cite{PLoSONE14,PCCP14}, 
and (v) the strength of the tether or linker connecting the motor and its cargo \cite{PCCP14}.
Anomalous active transport can also be typical for living cells \cite{PLoSONE14,PCCP14}, 
and indeed an increasing
number of experiments reveals its occurrence 
\cite{Seisenberg,Robert,Harrison}. Our modeling route is
based on non-Markovian  Generalized Langevin Equation (GLE) \cite{Kubo,Zwanzig}
description of viscoelasticity \cite{Mason,Waigh,Goychuk09,Goychuk12} and 
its multi-dimensional Markovian embedding 
\cite{Goychuk09,Goychuk12}. The approach is
deeply rooted in the main principles and dynamical foundation of statistical mechanics such as 
dynamical theory of Brownian motion and fluctuation-dissipation theorem \cite{Kubo,Zwanzig},
which must hold when the system is at thermal equilibrium. The approach explains the dynamical origin 
of both normal and anomalous Brownian motion and naturally extends beyond
thermal equilibrium, which makes it most suitable to describe physical
processes in living cells. Our theory naturally explains, in particular, why the power
exponent of anomalous active transport can be larger than one
of the passive subdiffusion and why the power exponent of anomalous active Brownian motion can be 
larger than doubled exponent of passive motion \cite{Robert,Harrison}. Such experimental
facts cannot be consistently 
explained within the previous approaches 
\cite{Caspi,BrunoPRE} to anomalous transport
by molecular motors, as detailed in \cite{PCCP14}. 
In fact, we develop a rather straightforward generalization
of the Brownian ratchets approach to modeling of molecular motors \cite{Chauwin,Julicher1,AstumianBier,
Astumian,Parmeggiani,Julicher2,Reimann,Makhno,Rosenbaum,Perez,Nelson}, with
a well proven utility in  the case of molecular motors underlying memoryless 
Markovian dynamics, towards anomalous non-Markovian dynamics with long-lasting memory 
reflecting viscoelastic effects in cytosol. Non-Markovian dynamics can
nevertheless  be considered as low-dimensional projection of a multi-dimensional 
Markovian dynamics -- the idea which
has a long tradition in statistical mechanics \cite{vanKampen}.

As it is is well known, microtubule is a periodic electrically polar structure 
featured by asymmetric periodic distribution of negative and positive charge densities on its surface \cite{Pollard,Baker}.
It has a spatial period $L=8$ nm. Furthermore,
the charge state of the motor protein depends on whether it is nucleotide
free (no extra charges are present), either ATP, or ADP and the phosphate group 
$\rm P_i$ are bound (three extra negative elementary  charges altogether in each case), or 
only ADP is bound (two extra negative elementary charges). 
Reflecting the change of charge distribution on the motor protein  the motor binding potential
in the electrical field of microtubule can also change accordingly \cite{Astumian}. 
Being periodic in space it should, however, 
be spatially asymmetric, and this asymmetry can direct molecular motor
in one selected direction (which can depend on the particular kind of motor) when the binding potential 
begins to stochastically fluctuate.  These fluctuations can be caused by changing the overall
charge of the motor protein accompanied by its conformational fluctuations 
due to the reactions of ATP binding, hydrolysis and dissociation 
of the products, repeated cyclically at random time instances. 
For two-headed kinesin molecules, one of the simplest, minimal model assumptions is
to assume just  two realizations  $U_{1,2}(x+L)=U_{1,2}(x)$ of the binding potential
$U(x,\zeta_i)$ depending on the motor state $\zeta_i$ with
the additional symmetry $U_{1,2}(x+L/2)=U_{2,1}(x)$ (the heads are assumed to be
identical and treated on equal footing) 
\cite{Julicher1,AstumianBier,Astumian,Parmeggiani,Julicher2,Makhno} . This ensures 
that two subsequent half-steps of the equal length $L/2$ makes 
one total step $L$ in the
direction defined by the asymmetry of the potential, 
see in Fig. \ref{Fig1}. Kinesins consume one ATP molecule 
per one full step.  
Hence, energy
$\Delta G_{\rm ATP}/2\approx 10\;k_BT_r\approx 0.25$ eV is invested into a half-step.
Simplest further assumption is to characterize the switching  process by one spatially-independent
rate $\nu_1=\nu_2$, so that the averaged motor turnover frequency is $\nu=\nu_1/2$.
A shortcoming of this approach is that it does not specify a mechanism of how the energy
of ATP hydrolysis is invested into the change of $U(x,\zeta_i)$. Neither reflects it the back
influence (i.e. the \textit{mutual} coupling) of the mechanical motion along microtubule on the
biochemical turnovers of the motor enzyme.
In particular, it is not consistent with the demand that if to gradually bring the reaction
of ATP hydrolysis to thermodynamic equilibrium, $\Delta G_{\rm ATP}\to 0$, the translation
motion of molecular motor must also gradually vanish.
Nevertheless, this ratchet approach remains immensely popular for Markovian dynamics
 \cite{Nelson,Reimann,AstumianBier,Makhno,Perez}. It
has been generalized for anomalously operating motors in Refs. \cite{PLoSONE14,PCCP14}. 
The role of mechanochemical coupling \cite{Nelson,Julicher1,Julicher2,AstumianBier,Fisher} is, however, 
important to address.
In the present work, we  consider a simplest popular model
of kinesins  with spatially- and $\Delta G_{\rm ATP}$-dependent transition 
rates $\nu_{1,2}(x)$  reflecting
a mechanochemical coupling of the translation motion of motor protein and its
conformational cyclic dynamics. We also consider an anharmonic model for tether connecting 
the motor and cargo by taking into account its maximally possible extension length, which
encompasses the previous harmonic model \cite{PCCP14} as a limiting case used for comparison.

\section{The model and methods}

\begin{figure}[t]
  \centering
  \includegraphics[width=7.5cm]{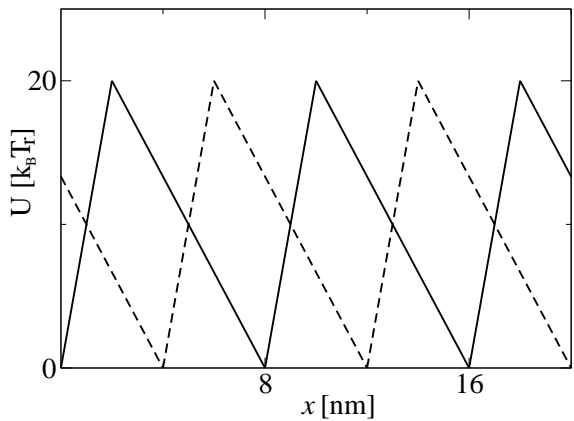}
 \caption{Typical sawtooth periodic binding potential reflecting spatial asymmetry of microtubule
 and its two realizations shifted by a half of the spatial period. }
  \label{Fig1}
\end{figure}

We start from considering subdiffusive overdamped 1d dynamics of cargo with radius $a_c$
and coordinate $y$ subjected to both viscous Stokes friction with friction coefficient 
$\eta_c$ and memory friction with kernel $\eta_{\rm mem}(t)$, as well as to unbiased Gaussian 
thermal random forces $\xi_c(t)$ and $\xi_{\rm mem}(t)$ of the environment at temperature $T$ which are
completely characterized by their autocorrelation functions
\begin{eqnarray}\label{1}
\langle \xi_c(t)\xi_c(t') \rangle = 2k_BT \eta_c\delta(t-t'),\\
\langle \xi_{\rm mem}(t)\xi_{\rm mem}(t') \rangle = k_BT \eta_{\rm mem}(|t-t'|).
\end{eqnarray}
The above relations express the second (classical) fluctuation-dissipation theorem (FDT) 
by Kubo \cite{Kubo,Zwanzig}. It has a very important physical content. Namely, at the thermal equilibrium 
the energy loss due to friction
is completely compensated by the energy gain from thermal stochastic force serving as
a ``stochastic lubricant''. Without it the motion would stop due to frictional losses. 
However, beyond thermal equilibrium there emerges an overall heat flux to
the environment, even in the absence of a temperature gradient.  
 Minimizing this flux one can arrive at the best thermodynamic efficiency of transport. 
This is a very important point: there is no need to minimize the friction, contrary to a popular belief,
but rather try to stay most closely to the thermal equilibrium, in order to minimize the heat
losses. Furthermore, the cargo is coupled to the motor with coordinate 
$x$ by an elastic tether or linker for which we use a 
finitely extensible non-linear elastic (FENE) model \cite{FENE} with coupling energy 
\begin{eqnarray}\label{FENE}
U_{mc}(r)=-\frac{1}{2}k_L r_{\rm max}^2 \ln \left ( 1- r^2/r_{\rm max}^2 \right ),
\end{eqnarray}
where $r=x-y$, and $k_L$ is elastic coupling constant. For a small extension, $r\ll r_{\rm max}  $,
$U_{mc}(r)\approx (1/2) k_L r^2$, recovering harmonic spring model, 
and the maximal extention length is $r_{\rm max}$.
The motor is also characterized by Stokes friction $\eta_m$ and the corresponding thermal
force $\xi_m(t)$ obeying FDT (\ref{1}) with $\eta_c\to\eta_m$. It moves in the binding
potential $U(x, \zeta(t))$ which depends on the motor conformation $\zeta(t)$. Altogether,
\begin{eqnarray}
\label{model1a}
\fl \eta_c \dot y &= &- \int_{0}^t\eta_{\rm mem}(t-t')\dot{y}(t')dt'-\frac{k_L(y-x)}{1-(y-x)^2/r_{\rm max}^2}
+\xi_c(t)+\xi_{\rm mem}(t),\\
\fl \eta_m\dot{x} &=&
\frac{k_L(y-x)}{1-(y-x)^2/r_{\rm max}^2}-\frac{\partial}{\partial x}U(x,\zeta(t))-f_0 +\xi_m(t),
\label{model1b}
\end{eqnarray}
where $f_0$ is an external loading force applied directly to the motor (for harmonic linker, 
$r_{\rm max}\to\infty$, it is the same as to apply it to the cargo). 

If the cargo is not coupled to the motor ($k_L\to 0$), 
the fractional memory
friction $\eta_{\rm mem}(t)=\eta_{\alpha}t^{-\alpha}/\Gamma(1-\alpha)$, $0<\alpha<1$, 
with fractional friction coefficient $\eta_{\alpha}$ leads to subdiffusion of cargo,  
$\langle\delta y^2(t) \rangle\approx 2D_\alpha t^\alpha/\Gamma(1+\alpha)$,
at the sufficiently large times $t\gg \tau_{\rm in}=(\eta_c/\eta_\alpha)^{1/(1-\alpha)}$.
Subdiffusion is characterized by the fractional diffusion coefficient $D_\alpha=k_BT/\eta_\alpha$.
Initially, for  $t\ll \tau_{\rm in}$ diffusion is, however, normal 
$\langle\delta y^2(t) \rangle\approx 2D_c t$ with $D_c=k_BT/\eta_c$. For all times
in this model \cite{KhG13a},
\begin{eqnarray}\label{exact}
\langle\delta y^2(t)\rangle=2D_ctE_{1-\alpha,2}\left(-[t/\tau_{\rm in}]^{1-\alpha}
\right),
\end{eqnarray}
where $E_{a,b}(z)=\sum_{n=0}^\infty z^n/\Gamma(an+b)$ is the generalized
Mittag-Leffler function.
For the coupled motor and cargo, 
in the limit of infinitely
rigid harmonic linker $k_L\to\infty$, $r_{\rm max}\to\infty$ 
one can exclude explicitely (see discussion in Ref. \cite{PCCP14}) the cargo dynamics and consider anomalous
dynamics of the motor alone, characterized by the Stokes friction $\eta_m+\eta_c$,
and the same memory friction.  This would provide a  
generalization of the model studied in Ref. \cite{PLoSONE14}. We consider here
but a more general model.

For the binding potential, we consider the same model of piecewise linear asymmetric 
potential with amplitude $U_0$, spatial period $L$, and the maximum dividing the spatial period
in the ratio $1:p$, with $p=3$, as in Ref. \cite{PCCP14}, see in Fig. \ref{Fig1}.
Similar models are well-known and widely used \cite{Astumian,Parmeggiani}.

\begin{figure}[t]
  \centering
  \includegraphics[width=10cm]{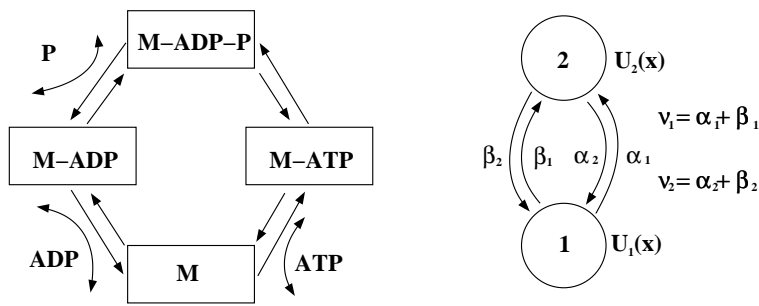}
 \caption{Biochemical cycle of the motor protein and its simplest two-state cycle reduction.}
  \label{Fig2}
\end{figure}

Important distinction between different further models possible occurs on the level of intrinsic enzyme
dynamics $\zeta(t)$. The chemical coordinate $\zeta(t)$ must be cyclic, with the cycle driven
preferably in one direction (rotation of ``catalytic wheel'' \cite{Wyman,Rosenbaum}) by the free 
energy $\Delta G_{\rm ATP}$ released from the reaction of ATP hydrolysis 
$\rm ATP\leftrightarrow ADP+P_i$ kept out of the thermodynamical equilibrium (shifted to the right)
by maintaining out-of-equilibrium concentrations of reactants. One of simplest related cycles involves
four discrete states (Fig. \ref{Fig2}, left) \cite{Julicher2}. 
The ratio of the product of the (pseudo-first order) rates 
(ATP, ADP, and $\rm P_i$ molecules are abundant with concentrations kept constant) of the
counter-clockwise transitions to the product of rates in the clockwise transitions must be equal
$\exp(\Delta G_{\rm ATP})/(k_BT)$ \cite{Hill,Hong}. In the case of two-headed kinesins, it 
should be applied to each head, with the rates which are position-dependent 
(mechano-chemical coupling) which would make already a rather 
complicate model. The simplest possible drastical
reduction accounting for the mechano-chemical coupling in the case of head-over-head motion
of kinesins, a so-called two-state cycle \cite{Hill,Julicher2,AstumianBier}, is shown in the 
right part of Fig. \ref{Fig2}. One considers only two states and two realizations
of $U_{1,2}(x)$ obeying additional symmetry $U_{1}(x+L/2)=U_{2}(x)$, with  each 
transition accounting for a half-step (one head moves, another remains bound). The corresponding
rates must be not only periodic in space, $\alpha_{1,2}(x+L)=\alpha_{1,2}(x)$, 
$\beta_{1,2}(x+L)=\beta_{1,2}(x)$, but also share additional symmetries, 
$\alpha_{1}(x+L/2)=\beta_{2}(x)$ and $\alpha_{2}(x+L/2)=\beta_{1}(x)$.
For a two-cycle it also must be:
\begin{eqnarray}
\frac{\alpha_1(x)\beta_2(x)}{\alpha_2(x)\beta_1(x)}=\exp(\Delta G_{\rm ATP}/(k_BT)),
\end{eqnarray}
for any $x$, which can be satisfied, e.g.,  by choosing
\begin{eqnarray}
\frac{\alpha_1(x)}{\alpha_2(x)}& = & \exp[(U_1(x)-U_2(x)+\Delta G_{\rm ATP}/2)/(k_BT)], \nonumber \\
\frac{\beta_1(x)}{\beta_2(x)}& = & \exp[(U_1(x)-U_2(x)-\Delta G_{\rm ATP}/2)/(k_BT)] .
\end{eqnarray}
The total rates 
\begin{eqnarray}
\nu_1(x)=\alpha_1(x)+\beta_1(x), \nonumber \\
\nu_2(x)=\alpha_2(x)+\beta_2(x)
\end{eqnarray}
of the transitions between two energy profiles must satisfy
\begin{eqnarray}
\frac{\nu_1(x)}{\nu_2(x)}=\exp[(U_1(x)-U_2(x))/(k_BT)]
\end{eqnarray}
at thermal equilibrium (thermal detailed balance condition of vanishing dissipative fluxes
both in the translation direction and within the conformational space of motor) 
\cite{Julicher1,AstumianBier}. 
It is obviously satisfied 
for $\Delta G_{\rm ATP}\to 0$, where both the flux in the transport direction and 
the flux within the chemical
coordinate space vanish together. 
Notice that if to choose $\nu_1=\nu_2=const$, the latter condition
is not possible to satisfy. 

We still have some freedom in choosing various models for $\alpha_1(x)$ 
or $\alpha_2(x)$. The rate $\alpha_1(x)$ corresponds to the reactions of ATP binding
and hydrolysis considered as one lump reaction. 
It is reasonable to assume that this rate is constant, $\alpha_1(x)=\alpha_1$ within some
$\pm\delta/2$ neighborhood of the minimum of potential $U_1(x)$ and is zero otherwise.
Correspondingly, the rate $\beta_2(x)=\alpha_1$ within 
$\pm\delta/2$ neighborhood of the minimum of potential $U_2(x)$. Given these assumptions we have: 
\begin{eqnarray}\label{nu1}
\fl \nu_1(x)& = &\alpha_1(x)+ \alpha_1(x+L/2)\exp[-(U_2(x)-U_1(x)
 + \Delta G_{\rm ATP}/2)/(k_BT)], \nonumber \\
\fl \nu_2(x)& = & \alpha_1(x)\exp[-(U_1(x)-U_2(x)+\Delta G_{\rm ATP}/2)/(k_BT)] 
 + \alpha_1(x+L/2)\;. \label{nu2}
\end{eqnarray}
Furthermore, if we choose $\delta=L/2$
(for the given model of binding potential with $p=3$),
then ATP binding to the motor and its 
hydrolysis can occur, in principle, anywhere on microtubule with the same rate. This is a reasonable
assumption from the biophysical point of view, which
lends a further support for our  model choice. 
It is easy to grasp that this model can give very similar results to the ratchet model with
spatially independent rates $\nu_1 = \nu_2 = \alpha_1$ \cite{PCCP14} 
for sufficiently large
$\Delta G_{\rm ATP}$ and potential amplitude $U_0$ having similar values. 
Then, $\nu\approx \alpha_1/2$ is approximately the motor turnover
frequency which is nearly independent of $x$.  This provides a possibility to compare the studied model
with the model in Ref. \cite{PCCP14}
by choosing other parameters appropriately.

\subsection{Energetics of the motor}

In Eqs. (\ref{model1a}), (\ref{model1b}), $R_m(t):=\eta_m\dot x(t)-\xi_m(t)$, 
$R_c(t):=\eta_m\dot y (t)-\xi_c(t)$, $R_{\rm mem}(t):=
\int_{0}^t\eta_{\rm mem}(t-t')\dot{y}(t')dt'-\xi_{\rm mem}(t)$ describe total 
environmental forces. The averaged work done by these forces,
 $\Delta Q_m(t)=\int_0^t \langle R_m(t')\dot x(t')\rangle dt'$,
$\Delta Q_c(t)=\int_0^t \langle R_c(t')\dot y(t') \rangle dt'$, 
$\Delta Q_{\rm mem}(t)=\int_0^t \langle R_{\rm mem}(t')\dot y(t') \rangle dt'$ is nothing else
as the total 
heat exchange with the environment, $\Delta Q(t)=\Delta Q_m(t)+\Delta Q_c(t)+\Delta Q_{\rm mem}(t)$.
At thermal equilibrium, $\lim_{t\to \infty}\Delta Q(t)/t=0$, i.e. overall heat exchange
is absent. Beyond thermal equilibrium, $\Delta Q(t)$ describes the heat transfer to the environment or
heat losses.  The averaged energy transduced by the potential flashes is 
$E_{\rm in,1}(t)=\int_0^t\left \langle  \partial U(x,\zeta(t'))/\partial t' \right\rangle dt'$ \cite{Sekimoto}, 
which yields the averaged sum of the binding potential jumps at the transition points $U_1\to U_2 \to U_1\to ...$. 
The mechanical work done by the motor against the external load $f_0$ is 
$W_{\rm use}(t)=f_0\Delta x(t)$. 
The energy balance is $E_{\rm in,1}(t)=\Delta Q(t)+W_{\rm use}(t)$, if to neglect
the back coupling of the potential fluctuations to the biochemical cycle of the motor, i.e.
the energy transferred back to the motor and the ATP energy source, which drives the whole
machinery.
Hence, thermodynamic efficiency within such a treatment is 
\begin{eqnarray}\label{eff1}
R_{\rm th,1}(t)=\frac{W_{\rm use}(t)}{E_{\rm in,1}(t)}.
\end{eqnarray}
It becomes constant for a sufficiently large $t$, in the case of normal transport, where
both $E_{\rm in,1}(t)\propto t$, and $W_{\rm use}(t)\propto t$, but algebraically decays
to zero, $R_{\rm th,1}(t)\propto 1/t^{1-\alpha_{\rm eff}}$, in the case of anomalous transport,
$\Delta x(t)\propto t^{\alpha_{\rm eff}}$, in the cases of a periodic
potential modulation \cite{GKh13a,KhG13a}, or constant potential-independent flashing rates 
\cite{PLoSONE14,PCCP14}. 
In the absence of external load, 
$f_0=0$, $R_{\rm th,1}(t)=0$,
i.e. all the input energy is eventually dissipated as heat. However, something useful
is yet done. Namely, the cargo is
 transferred on some distance $d(t)$. Different Stokes efficiencies have been defined to characterize
energetic performance of motors in such a situation for memoryless friction \cite{Derenyi,Suzuki,WangOster}. 
However, the notion
of Stokes efficiency becomes even more ambiguous for viscoelastic environment \cite{KhG13a}. 
For this reason,
a delivery efficiency has been introduced in Ref. \cite{PLoSONE14}. It is the ratio of the mean 
velocity $v(t)=d(t)/t$ of the cargo delivery to the mean number $\langle N_{\rm turn}(t)\rangle$ 
of motor enzyme turnovers made, 
\begin{eqnarray}\label{delivery_eff}
{\rm D}=\frac{d}{t \langle N_{\rm turn}\rangle}.
\end{eqnarray}

The definition (\ref{eff1}) is the only possibility to define thermodynamic efficiency
in the case of ratchet models  which do not specify the mechanism 
of mechanochemical coupling. 
Within these models, $\zeta(t)$ is considered as a driver which provides input energy 
\textit{unidirectionally}, i.e. without feeling any feedback \cite{Sekimoto}.
However, within the considered model with a \textit{mutual}
coupling of the conformational cycling of the motor enzyme and its translational motion
the interaction energy
$U(x,\zeta(t))$ provides a bidirectional coupling. The energy can flow in both directions. 
The energy supply is provided by a pool of ATP molecules which is characterized
by out-of-equilibrium chemical  potential difference of the reaction 
of ATP hydrolysis, 
which we denoted as $\Delta G_{\rm ATP}$.
Hence, it is reasonable to define
the input energy as 
$E_{\rm in,2}(t)=\Delta G_{\rm ATP}\langle N_{\rm turn}(t)\rangle$ \cite{Julicher1}. 
Then, the thermodynamic
efficiency is 
\begin{eqnarray}\label{eff2}
R_{\rm th,2}(t)=\frac{W_{\rm use}(t)}{\Delta G_{\rm ATP}\langle N_{\rm turn}(t)\rangle}.
\end{eqnarray} 
The idea is clear: each turnover of the catalytic wheel requires energy of the hydrolysis of
one ATP molecule, in accordance with the main principles of the non-equilibrium thermodynamics
applied to the biochemical cycle kinetics \cite{Hill,Hong}. One expects that
$R_{\rm th,2}(t)>R_{\rm th,1}(t)$. This is because a part of the energy $E_{\rm in,1}(t)$
can be given back to the motor, its intrinsic degree of freedom, and e.g. 
recuperated in the backward synthesis, $\rm ADP+P_i\rightarrow ATP$. 
This feature is beyond simple ratchet models with spatially independent rates 
which do not take properly into account such a mechano-chemical coupling.
However, this definition is also not quite precise. As a matter of fact, 
one ATP molecule is only consumed if a cycle is accomplished in the counter-clockwise 
direction in the 
left diagram of Fig. \ref{Fig2}. Moreover, it is recuperated
if the cycle is completed in the clockwise direction. Therefore, for the two-state
cycle depicted in the right diagram of Fig. \ref{Fig2}, to correctly calculate consumption
of ATP molecules we should count $p_1\Delta G_{\rm ATP}/2$, with 
$p_1=(\alpha_1-\beta_1)/(\alpha_1+\beta_1)$ for the transition $U_1\to U_2$, and 
$p_2\Delta G_{\rm ATP}/2$
with  $p_2=(\beta_2-\alpha_2)/(\alpha_2+\beta_2)$ for the transition $U_2\to U_1$. 
The correspondingly calculated
input energy is denoted as $E_{\rm in,3}(t)$, and thermodynamic efficiency as $R_{\rm th,3}(t)$.
Obviously, since $E_{\rm in,3}(t)<E_{\rm in,2}(t)$, $R_{\rm th,3}(t)>R_{\rm th,2}(t)>R_{\rm th,1}(t)$.
However, in a regime, where the catalytic wheel rotates 
overwhelmingly in one direction (like in Michaelis-Menthen treatment of enzymatic
reactions, where the backward rotation is entirely neglected), 
the distinction between $R_{\rm th,3}(t)$ and 
$R_{\rm th,2}(t)$ becomes negligible. By the same token, 
one can modify the definition of the delivery efficiency in Eq. (\ref{delivery_eff})
by replacing $\langle N_{\rm turn}\rangle)$ with the averaged number of ATP molecules
hydrolyzed. However, the difference between both definitions exists only beyond
the Michaelis-Menthen description of the motor kinetics.

\subsection{Markovian embedding and numerical method}

Following to already well-established methodology of Markovian embedding of Refs. 
\cite{Goychuk09,Goychuk10,Goychuk12}
we approximate power-law memory kernel $\eta_{\rm mem}(t)$ by a sum of exponentials, 
\begin{eqnarray}\label{kernel2}
\eta_{\rm mem}(t)=\sum_{i=1}^N k_i \exp(-\nu_i t),
\end{eqnarray}
obeying fractal scaling $\nu_i=\nu_0/b^{i-1}$, $k_i\propto \nu_i^\alpha$, 
and introduce $N$ auxiliary Brownian particles modeling
viscoelastic properties of the environment. This allows to transform the considered 
non-Markovian problem into a Markovian problem in the space of enhanced (by $N$) 
dimensionality. Then, the standard methods of integration stochastic differential
equations (SDEs), such as stochastic Euler, or stochastic Heun method,
can be applied for a fixed realization of the potential $U_i(x)$. 
The method allows for a highly accurate numerical integration of fractional
Langevin dynamics even for a moderate $N\sim 10-100$ \cite{Goychuk09}. The accuracy of kernel approximation
is controlled by the scaling parameter $b$ and even for the decade scaling 
$b=10$ expressing the idea ``one power law time decade requires about 
one exponential in doing approximation'' it is better than 4\% between two memory cutoffs,
short time cutoff $\tau_{\rm min}=1/\nu_0$, and large time  cutoff
$\tau_{\rm max}=b^{N-1}\tau_{\rm min}$. For $b=2$, the accuracy of approximation
improves to about $0.01\%$ \cite{GKh13a}. Statistical errors in numerical simulations due to a finite
number of trajectory realizations $n$ will
always be larger for a practical $n$, scaling
as $1/\sqrt{n}$. For $n\sim 10^3-10^4$, $b=10$ suffices for most practical purposes
with the accuracy of several percents which in stochastic simulations is
considered as a very good one.
Given the maximal time range of subdiffusion defined by 
$\tau_{\rm max}$, one can always find appropriate minimal Markovian embedding
with the required accuracy. Since $N\sim \log_b (\tau_{\rm max}/\tau_{\rm min})$
this ensures excellent numerical method \cite{Goychuk09,Goychuk12}. The finiteness of $\tau_{\rm max}$
reflects finite effective viscosity $\zeta_{\rm eff}$  of cytosol fluid which can be exponentially
enhanced with respect to one of water depending on the cargo size \cite{Odijk,Masaro,Holyst}. Indeed,
the effective friction at very large times is 
$\eta_{\rm eff}=\int_0^{\infty}\eta_{\rm mem}(t)dt \propto \zeta_{\rm eff}$ and
on the scaling grounds, $\eta_{\alpha}\sim \eta_{\rm eff}\tau_{\rm max}^{\alpha-1}$.
For $t\gg \tau_{\rm max}$, passive diffusion of cargo becomes again normal. It is characterized
by the friction coefficient $\eta_c+\eta_{\rm eff}$ and diffusion coefficient 
$D_{c,\rm eff}=k_BT/(\eta_c+\eta_{\rm eff})$ largely suppressed with respect to water.
With these  parameters \cite{PLoSONE14},
\begin{eqnarray}
k_i=\nu_0\eta_{\rm eff}\frac{b^{1-\alpha}-1}{b^{(i-1)\alpha}[b^{N(1-\alpha)}-1]}.
\end{eqnarray}

Following to \cite{Goychuk12,PCCP14}, upon introduction of $N$ auxiliary overdamped Brownian particles
with coordinates $y_i$ and frictional coefficients $\eta_i=k_i/\nu_i$, the Markovian 
embedding dynamics reads
\begin{eqnarray}
\label{embedding}
\eta_m\dot{x}&=&f(x,\zeta(t))+\frac{k_L(y-x)}{1-(y-x)^2/r_{\rm max}^2}+\sqrt{2\eta_m k_BT}\xi_m(t),\nonumber\\
\eta_c\dot{y}&=& -\frac{k_L(y-x)}{1-(y-x)^2/r_{\rm max}^2}-\sum_{i=1}^{N}k_i(y-y_i)+\sqrt{2\eta_c k_BT}\xi_0(t),
\nonumber\\
\eta_i\dot{y_i}&=&k_i(y-y_i)+\sqrt{2\eta_ik_BT}\xi_i(t),
\end{eqnarray}
where $f(x,\zeta(t))=-\partial U(x,\zeta(t))/\partial x-f_0$, and $\eta_i=k_i/
\nu_i$. Furthermore,  $\xi_i(t)$ are uncorrelated white Gaussian noises of
unit intensity, $\langle \xi_i(t')\xi_j(t)\rangle=\delta_{ij}\delta(t-t')$, which
are also uncorrelated with the white Gaussian noise sources $\xi_0(t)$ and $\xi_m(t)$. 
To have a complete equivalence with the stated GLE model in Eqs.~(\ref{model1a}), (\ref{model1b})
with memory kernel (\ref{kernel2}), the initial positions $y_i(0)$ are sampled from a Gaussian
distribution centered around $y(0)$, $\langle y_i(0)\rangle=y(0)$ with variances
$\langle[y_i(0)-y(0)]^2\rangle=k_BT/k_i$ \cite{Goychuk12}.

\subsubsection{Choice of parameters and details of numerics.}

As in Ref. \cite{PCCP14}, we take  $a_m=100$ nm for the effective radius of kinesin,
about 10 times larger than its linear geometrical size (without tether) in order to account for the
enhanced effective viscosity experienced by the motor in the cytosol compared
to its value in water. The viscous friction coefficient is estimated from the Stokes formula
as $\eta_m=6\pi a_m\zeta_w$, where $\zeta_w=1\;{\rm mPa\cdot s}$ is water viscosity.
Furthermore, we use the characteristic time scale $\tau_m=L^2\eta_m/U_0^
*$ to scale time in the numerical simulations with $U_0^*=10\;k_BT_r$. For the
above parameters, $\tau_m\approx 2.94\;\mu{\rm s}$. Distance is scaled in units
of $L$,  elastic coupling constants in units of $U_0^*/L^2\approx 0.64$ pN/nm, and forces
in units of $U_0^*/L\approx 5.12$ pN. $\nu_0$ was chosen $\nu_0=100$ ($3.4\cdot 10^7$ 1/s)
yielding $\tau_{\rm min}=29.4$ ns,  
and $\alpha$ was $\alpha=0.4$ as found experimentally in \cite{Robert,Bruno}. Two cargo sizes were considered, large $a_c=300$ nm, which corresponds
to the magnetosome size in Ref. \cite{Robert}, and a smaller one. For larger cargo, we assume
that its effective Stokes friction $\eta_c=6\pi a_c\zeta_w$ is enhanced by the factor 
of $\eta_{\rm eff}/\eta_c=3\cdot 10^4$ in cytosol. Assuming that $\tau_{\rm max}=10^9\tau_{\rm min}=29.4$ s
this yields fractional friction coefficient 
$\eta_{\alpha}= \eta_{\rm eff}\tau_{\rm max}^{\alpha-1}/r$ with $r\approx 0.93$ \cite{PCCP14},
which yields subdiffusion coefficient 
$D_{0.4}=k_BT/\eta_{0.4}\sim 1.71 \cdot 10^{-16}\;{\rm m^2/s^{0.4}}=171 \;{\rm nm^2/s^{0.4}}$.
It is in a semi-quantitative agreement with the experimental results in \cite{Robert}.
Smaller cargo is characterized by $\eta_{\rm eff}/\eta_c=3\cdot 10^3$ yielding 
$D_{0.4}=1710 \;{\rm nm^2/s^{0.4}}$, ten times larger. Furthermore, we used two values
of rate constant $\alpha_1$: $170\;\rm{s}^{-1}$ (fast) and $34\;\rm{s}^{-1}$ (slow),
in order to match approximately the enzyme turnover rates $\nu \sim \alpha_1/2$
in Ref. \cite{PCCP14}. Accordingly, we used mostly $U_0=20$ $k_BT_{ r}$ in simulations, although
we used also two larger values of $U_0$, see Table \ref{Table1}, in order to arrive
at the thermodynamical efficiencies as large as 50\% typical for kinesins \cite{Nelson}. 
Moreover, two different values for the elastic spring constant were used, $k_L^{(1)}=0.32$
pN/nm (`strong'), which corresponds to measurements in vitro \cite{Kojima}, and
a ten times softer spring $k_L^{(2)}=0.032$ pN/nm (`weak'), in accordance with
recent results in Ref.~\cite{Bruno} in living cells. For the maximal extension of linker
we used $r_{\rm rmax}=80$ nm \cite{Pollard,Hirokawa}, and also $r_{\rm rmax}=\infty$, which corresponds to harmonic
linker in Ref. \cite{PCCP14}. The studied set of parameters is shown in Table \ref{Table1}. 

\begin{table} 
\begin{center}
\caption{Parameter sets}\label{Table1}\vspace{0.5cm}

\begin{tabular}{|p{0.75cm}|p{1.4cm}|p{1.3cm}| p{0.8cm}|p{0.8 cm}| p{1.0cm}|}
\hline
Set & $D_{0.4}$, ${\rm nm^2/s^{0.4}}$ & $k_L$, pN/nm & $\alpha_1$, $\rm{s}^{-1}$ & $U_0$, $k_BT_{ r}$  
& $r_{\rm rmax}$, nm \\
\hline
$S_{1}$ &  171 & 0.320 & 170 & 20 & $\infty$ \\ 
$S_2$ &  1710 & 0.320 & 170 & 20 & $\infty$ \\
$S_3$ &  1710 & 0.032 & 170 & 20 & $\infty$ \\ 
$S_{4a}$ &  171 & 0.032 & 170 & 20 & $\infty$ \\
$S_{4b}$ &  171 & 0.032 & 170 & 20 & 80\\
$S_5$ &  171 & 0.320 & 34 & 20 & $\infty$ \\
$S_6$ &  171 & 0.032 & 34 & 20 & 80 \\
$S_7$ &  171 & 0.320 & 170 & 25 & 80 \\
$S_8$ &  171 & 0.320 & 170 & 30 & 80 \\
$S_9$ &  1710 & 0.320 & 170 & 25 & 80 \\
$S_{10}$ &  1710 & 0.320 & 170 & 30 & 80 \\
\hline
\end{tabular}
\end{center}
\end{table}

In order to numerically integrate stochastic Langevin dynamics following
to one potential realization $U_{1,2}(x)$, we used stochastic Heun method 
implemented in parallel on NVIDIA Kepler graphical processors.
Stochastic switching between two potential realizations is realized using
a well-known algorithm. Namely, if the motor is on the given surface $U_1(x)$ or $U_2(x)$, 
at each integration  time step it can switch with the 
probability $\nu_{1}(x)\delta t$ or $\nu_{2}(x)\delta t$ to another surface, or to
evolve further on the same surface, where $\delta t$ is the integration time step,
and $\nu_{1,2}(x)$ are given in Eq. (\ref{nu1}).
A particular embedding 
with $b=10$ and $N=10$ was chosen in accordance with our previous studies. Furthermore,
we use  $\delta t=5\cdot 10^{-3}$ and $n=10^3$ trajectories used for the ensemble averaging.
The maximal time range of integration was $10^6$, which corresponds to $2.94$ sec 
of motor operation. $\Delta G_{\rm ATP}=20$ $k_BT_r$ was taken in all numerical simulations.
We also checked that with $\Delta G_{\rm ATP}\to 0$, the motor stops at $f_0=0$, i.e. 
no directed motion and useful work can be derived from the reaction of ATP hydrolysis
being at thermal equilibrium, in
accordance with stochastic thermodynamics of isothermal engines \cite{Hill,Hong}.

\section{Results and Discussion}

\begin{figure}[t]
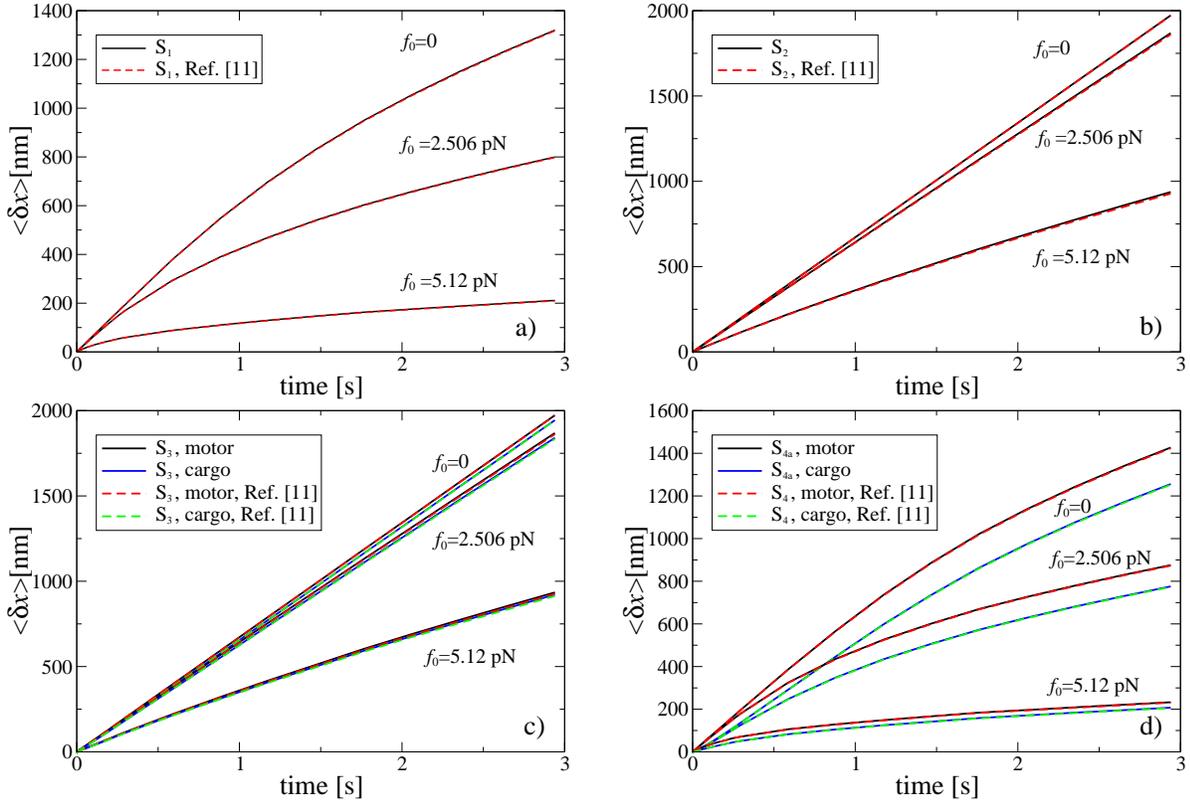

  \includegraphics[width=7.5cm]{Fig3a.eps} \hfill
   \includegraphics[width=7.5cm]{Fig3b.eps} \vfill
   \includegraphics[width=7.5cm]{Fig3c.eps} \hfill
    \includegraphics[width=7.5cm]{Fig3d.eps}
 \caption{(a) Mean distance traveled by motor for the set of parameters $S_{1}$ (black continuous lines) 
 versus time is compared with one for the set of parameters $S_1$ in Ref. \cite{PCCP14} (red dashed lines),
 for three different values of the opposing external force $f_0$ acting on the motor: $0$, $2.506$ pN,
 and $5.12$ pN (from top to bottom). (b) The same for the set $S_2$ here and the 
 set $S_2$ in Ref. \cite{PCCP14}.
 (c) The same for the set $S_3$ here and the set $S_3$ in Ref. \cite{PCCP14}.
 (d) The same for the set $S_{4a}$ here and the set $S_4$ in Ref. \cite{PCCP14}.
 In (c), (d), which correspond to weaker linker, we also show the mean position of cargo 
 (full blue lines for the present model, and dashed orange lines for the model in Ref. \cite{PCCP14}).
 It is seen, that for larger cargo in (d) the distance between the motor and cargo grows
 in time, which will result in the disruption of such a weak harmonic linker \cite{PCCP14}.
 This is where the linker anharmonicity (maximal extension possible) can play indeed  a very important
 role.
 }
  \label{Fig3}
\end{figure}

\begin{figure}[t]
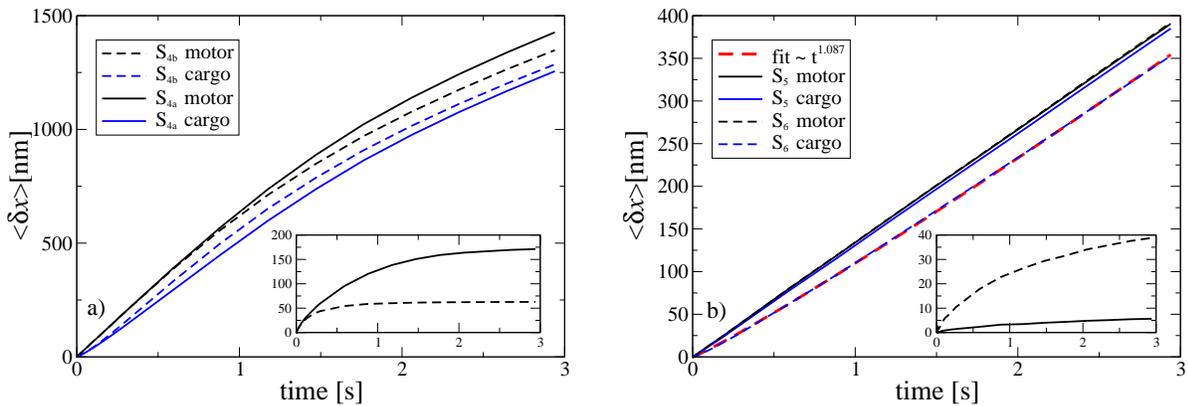

  \includegraphics[width=7.5cm]{Fig4a.eps} \hfill
  \includegraphics[width=7.5cm]{Fig4b.eps} 
 \caption{ (a) Mean motor (black) and cargo (blue) positions for the sets $S_{4a}$ (full lines)
 and $S_{4b}$ (dashed lines) and their differences in the inset. External load $f_0=0$.
 (b) The same as (a) for the sets $S_5$ (full lines) and $S_6$ (dashed lines).
 }
  \label{Fig4}
\end{figure}

We compare first in Fig. \ref{Fig3} the results for the parameter sets $S_{1}$, $S_2$, $S_3$, $S_{4a}$ 
of the present model and the 
corresponding parameter sets $S_1$, $S_2$, $S_3$, $S_4$
in Ref. \cite{PCCP14} , for the ensemble averaged trajectories.
One can see that almost no difference can be visually detected, both for larger and
smaller cargo, stronger and weaker linker. The transport
is clearly anomalously slow for $S_{1}$ and $S_{4a}$ always (large cargo), and it changes from normal
to anomalous transport upon  increase of $f_0$ in the cases $S_2$ and $S_3$ (smaller cargo). For 
stronger
linker, the difference between the motor and cargo positions is not significant, and for this reason
the cargo position is not shown in Fig. \ref{Fig3}, (a),(b). For weaker linker, the
difference becomes very strong in the case
of large cargo, see in Fig. \ref{Fig3}, d, where for $f_0=0$ it increases up to about $170$ nm,
see in the Fig. \ref{Fig4}, a, inset.
The corresponding elastic energy becomes $111$ $k_BT_{r}$, i.e. of 
the same of order as a typical energy of covalent
bonds and such a linker clearly cannot sustain transport \cite{PCCP14}. It will be disrupted.
However, the linker anharmonicity starts to play a profound role when the cargo-motor distance
becomes larger than about one fifth of the maximal linker extension $r_{\rm max}$. The latter one
can be in the range between 10 nm and 150 nm for different motors \cite{Pollard,Hirokawa}. 
We consider $r_{\rm max}=80$ nm
($10\;L$). Then, for a strong linker the anharmonicity does not play any essential role.
We clarify its role in Fig. \ref{Fig4}, a for a weak linker and large cargo. It is seen that 
anharmonicity restricts the increase of the cargo-motor distance by $r\approx 63$ nm. Substituting
this value in Eq. (\ref{FENE}) yields $U_{mc}^{(\rm max)}\approx 24$ $k_BT_{ r}$.
Such a linker should be able to sustain transport of large cargo,  not necessarily it will be 
disrupted. This is an  important result:
Even weak linkers can possibly support strongly anomalous transport of large cargos due to 
nonlinear effects.

In Ref. \cite{PCCP14}, we revealed a very interesting effect which can emerge due to the weakness of
linker. Namely, if to reduce the turnover frequency of motor pulling large cargo from 85 to $17$ Hz
(which is the case $S_6$ in  \cite{PCCP14})
then the motor operates normally at $f_0=0$, whereas the cargo enters temporally a super-transport 
regime with $\alpha_{\rm eff}>1$. A natural question emerges if this effect survives for the considered 
FENE model of linker. Fig. \ref{Fig4}, b answers this question in affirmative, lending it therefore
further support with respect to possible experimental verification. 
The explanation of this effect is simple: 
When the motor is in normal regime, its distance increases linearly in time. However, for a weak
linker the retardation of the cargo past the motor increases sublinearly in time, see inset
in Fig. \ref{Fig4}, b. This causes the effect that the mean distance covered by
cargo increases super-linearly in time, although it, in fact, moves slower than motor.
Clearly, in this case ``super''-transport does not imply a faster transport at all! The cargo lags behind
the normally walking motor. Interestingly, sub-transport also does not
proceed necessarily slower than the normal one \cite{GoychukFNL,GoychukPRE12,PRL14}.

\begin{figure}[t]
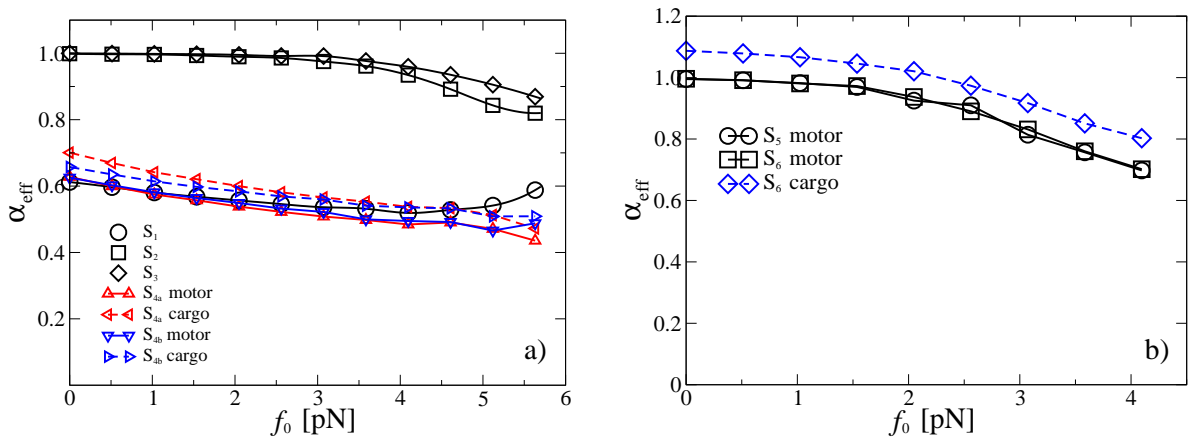

  \includegraphics[width=7.5cm]{Fig5a.eps} \hfill
  \includegraphics[width=7.5cm]{Fig5b.eps} 
 \caption{ Dependence of the effective transport exponent $\alpha_{\rm eff}$ on $f_0$
 for motor and cargo for different sets of the parameters indicated
 in plots.
 }
  \label{Fig5}
\end{figure}

The effective transport exponents $\alpha_{\rm eff}$ of motor and cargo (the latter 
one in some cases only, 
where the difference is substantial)
are shown in Fig. \ref{Fig5}. Their behavior is rather similar to one studied in \cite{PCCP14}.
The new feature is that the difference between $\alpha_{\rm eff}$ for the motor and cargo
in the case of the transport of large cargo on weaker linker becomes smaller due to 
nonlinear effects in elastic coupling. This is, however, what was to expect, not a surprise.

\subsection{Thermodynamic efficiency}

\begin{figure}[t]
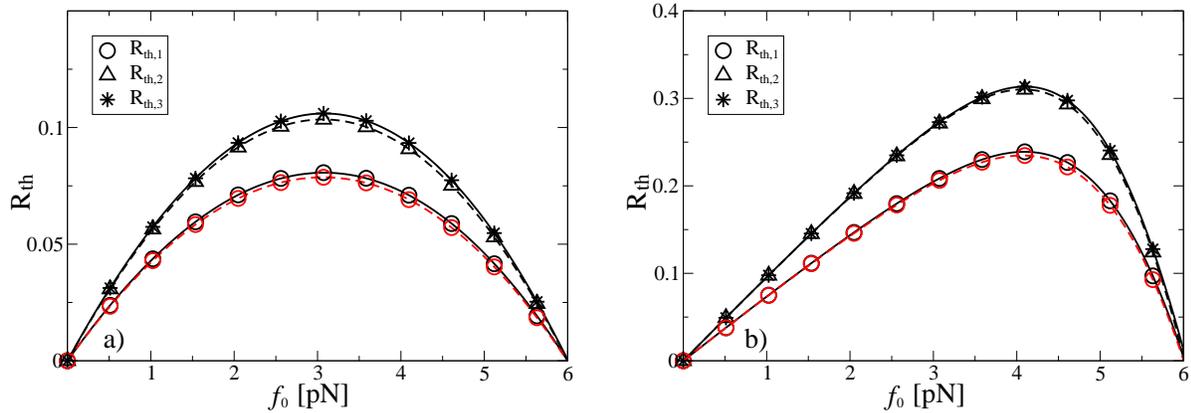

  \includegraphics[width=7.5cm]{Fig6a.eps}  \hfill
  \includegraphics[width=7.5cm]{Fig6b.eps} 
 \caption{ Differently defined thermodynamic efficiencies for the sets $S_1$ (a)
 and $S_2$ (b) at the end point
 of simulations. The red symbols connected by dashed red lines
 show the results for the sets $S_1$ and $S_2$ from Ref. \cite{PCCP14}.
 }
  \label{Fig6}
\end{figure}

The real discrepancies between the studied model and the model in \cite{PCCP14}
appears only for the thermodynamic efficiency,
see in Fig. \ref{Fig6}. Indeed, $R_{\rm th,2}$ is essentially larger than 
$R_{\rm th,1}$, and $R_{\rm th,3}$ is slightly larger than $R_{\rm th,2}$.
The latter relative discrepancy is, however, less than 2\% for the set $S_1$
(strongly anomalous transport) and becomes almost negligible for the set 
$S_2$ (close to normal transport),
indicating that ``catalytic wheel'' rotates overwhelmingly in one direction. 
Notice also that the difference between $R_{\rm th,1}$ and the same quantity
for similar parameter sets in Ref. \cite{PCCP14} is small. This once more
confirms that the ratchet models with constant, spatially independent rates
provide a reasonable description of the work of molecular motors. What they, however,
cannot do properly indeed 
is to describe thermodynamic efficiency of the motor.
This is a principal shortcoming because simple ratchet models do not take
properly the (bidirectional) mechano-chemical coupling into account.
The correct definition of thermodynamic efficiency of molecular motors is one 
given by $R_{\rm th,3}$, and it can be essentially larger than $R_{\rm th,1}$.

However, the normal \textit{modus operandi} of linear molecular motors such as 
kinesins is one at zero thermodynamic efficiency ($f_0=0$). The work is done
entirely on overcoming the dissipative resistance of the environment while
relocating cargo from one place in the cell to another one. Neither potential energy
of the motor, nor the potential energy of the cargo is enhanced at the end.
This is very different from the work of the ionic pumps whose primary goal
is to enhance the electrochemical potential of the pumped ions.
 The cargo 
delivery efficiency $\rm D$ exhibits the same features revealed in Refs. 
\cite{PLoSONE14,PCCP14}, and we do not consider it in further detail in this paper,
referring the readers to Ref. \cite{PCCP14}.
As a matter of fact, all the main features revealed in Refs. \cite{PLoSONE14,PCCP14} with respect
to occurrence of normal \textit{vs.} anomalous transport regime depending especially
on the cargo size, binding potential amplitude, and motor operating frequency
remain valid, being even rather close in numerical values, if to match
the parameters of both models appropriately.
This confirms that the modeling route of flashing ratchets with spatially independent
rates is a very reasonable one. 

\subsection{Anomalously slow motor turnovers with high thermodynamic efficiency}

\begin{figure}[t]
  \centering
  \includegraphics[width=10cm]{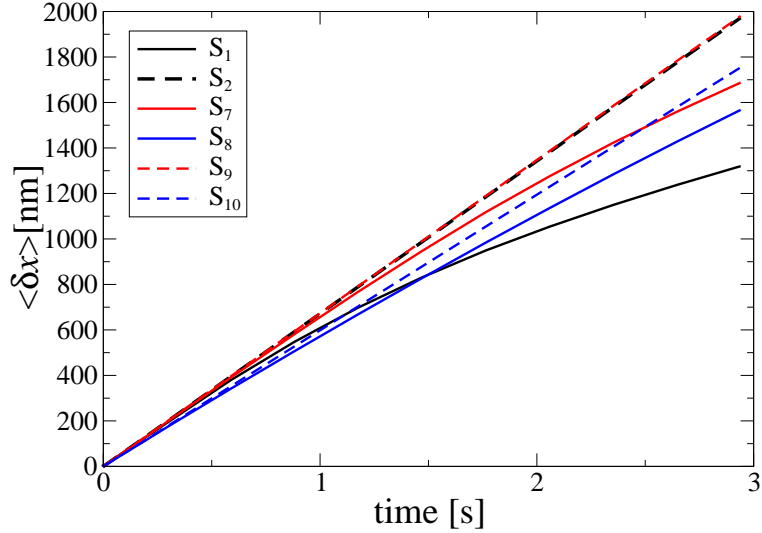}  
 \caption{Influence of the potential height $U_0$ on transport
 of cargos for different sets of parameters:  $S_1$, $S_7$, $S_8$ (large cargo), 
 and $S_2$, $S_9$, $S_{10}$ (smaller cargo), which differs only by $U_0$, at 
 zero $f_0=0$. For $U_0=25$  the transport
 becomes faster in absolute terms,  for large cargo, see set $S_7$,
 or proceeds with the same optimal speed, $v=L\nu_{\rm turn}$ for smaller cargo, set $S_9$,
 reflecting perfect synchronization of the enzyme turnovers with its translational
 motion, see in Ref. \cite{PCCP14}. 
 However, it is getting slower  again with a
 further increase of $U_0$ (sets $S_8$ and $S_{10}$).
 }
  \label{Fig7}
\end{figure}

The last question, which we shall clarify, is that whether this simple model can
demonstrate thermodynamic efficiency as high as 50\% featuring real kinesins with
stalling force about 7-8 pN \cite{Nelson}. Indeed, this is the case. As a guiding consideration let 
us start from the expression
for the stalling force obtain in Ref. \cite{PCCP14} by fitting numerical simulations
therein:
\begin{eqnarray}\label{stop_force}
f_0^{\mathrm{stall}}(T,U_0,\nu_{\rm turn})\approx\frac{4}{3L}F_0(T,U_0,\nu_{\rm turn}),
\end{eqnarray}
for $F_0>0$ with $F_0(T,U_0,\nu_{\rm turn})=U_0-U_m(\nu_{\rm turn})T/T_r:=U_0-TS_0(\nu_{\rm turn})$, 
and $U_m\approx 11.2\;k_BT_r$ at $\nu_{\rm turn}=85$
Hz, or $\alpha_1=170$ $\rm s^{-1}$, in our case. $F_0(T,U_0,\nu_{\rm turn})$ can be interpreted
as free-energy barrier height. At $T=0$, $f_0^{\mathrm{stall}}=4U_0/(3L)$, the result
which is easy to obtain due to the piece-wise constant character of the force
in the considered binding potential. Temperature reduces $F_0$ due to entropic contribution
$S_0$, and for $F_0<0$,
the stalling force is exponentially small. This imposes the condition of minimal $U_0$
for molecular motors at physiological temperatures being  $10-11$ $k_BT_r$ \cite{PCCP14}.
Eq. (\ref{stop_force}) yields $f_0^{\mathrm{stall}}\approx 6$ pN, 
at $U_0=20$ $k_BT_r$ and $\alpha_1=170$ $s^{-1}$, in agreement with numerics. It also predicts 
$f_0^{\mathrm{stall}}\approx 9.43$ pN at $U_0=25$ $k_BT_r$, and 
$f_0^{\mathrm{stall}}\approx 12.85$ pN at $U_0=30$ $k_BT_r$.
This suggests to use $U_0$
in the range of $20-30$ $k_BT_r$
to describe a realistically strong motor with larger efficiency. 
Simple ratchet models, which do not take properly the mechanochemical coupling into account,
 may prevent the detailed consideration of such high binding potential
amplitudes because they create impression that the energy of the hydrolysis of one 
ATP molecule may simply be not enough to fuel one catalytic cycle and to move synchronously 
by one spatial period at the same time. This is 
because the sum of  the energies required to lift the potential energy of the
motor in the binding potential while doing two half-steps becomes larger than $\Delta G_{\rm ATP}$.
Such an argumentation, however, neglects the fact the energy invested in the enhancement of the
motor's potential
energy can be recuperated and used again. In fact, even for $U_0=30$ $k_BT_{ r}$ (sets $S_8$, $S_{10}$)
the motor moves remarkably fast, faster than for $U_0=20$ $k_BT_{r}$,
in absolute terms, at the end point of simulations, but yet slower for intermediate times (set $S_8$),
see in Fig. \ref{Fig7}.

However, the motor becomes slower for $U_0=30$ $k_BT_{ r}$  than
for $U_0=25$ $k_BT_{ r}$ (sets $S_7$, $S_{9}$) at the same other parameters.
This slowdown results obviously because the motor turnovers
became slower. The stalling force about $10$ pN at $U_0=30$ $k_BT_{ r}$ is essentially lower 
than one predicted by Eq. (\ref{stop_force}). This is because the motor operation frequency
is lower than $85$ Hz. More precisely, it cannot be characterized  by a turnover frequency anymore,
at least when it pulls a large cargo.
Our analysis, see below, reveals that in this case the input energy $E_{\rm in,3}$ grows sublinearly
in time,  $E_{\rm in,3}(t)\propto t^{\gamma} $, $0<\gamma<1$, or 
$\langle N_{\rm turn}(t)\rangle \propto t^{\gamma} $, meaning that the enzyme turnovers
become anomalously slow, and it cannot be characterized anymore by a frequency. This
is a profoundly new  result: The mechano-chemical coupling can cause anomalously
slow rotation of the catalytic wheel. Such an enzymatic reaction cannot be characterized
by a mean rate anymore!
The intuition is correct in predicting that it will be difficult 
to rotate one enzymatic cycle using energy of one ATP molecule for such a large $U_0$. 
However, the motor operation is still possible, and it can start to consume ATP energy
\textit{sublinearly} in time, with exponent $\gamma$.
Astonishingly, the motor can become thermodynamically highly efficient in this anomalous
regime, see below. 

\begin{figure}[t]
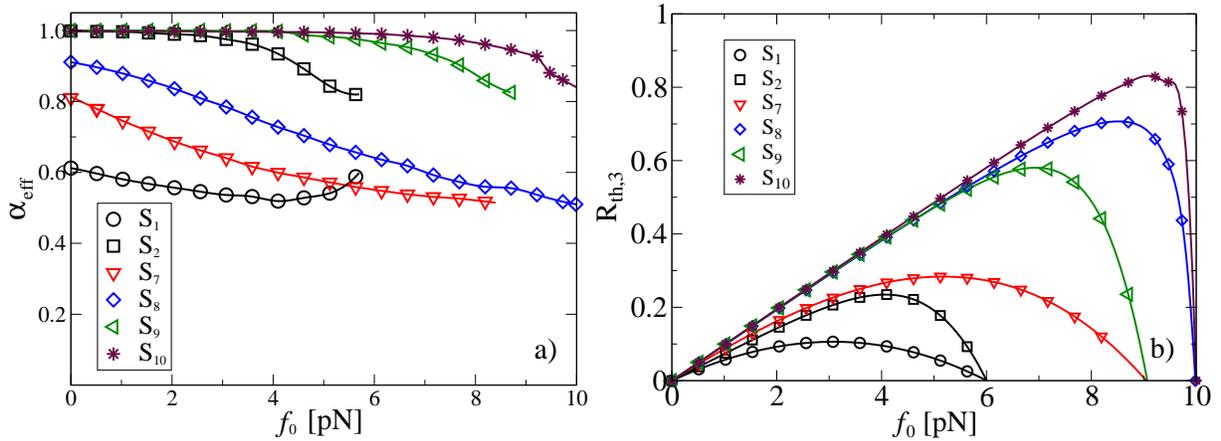

\vspace{0.5cm}
  \includegraphics[height=5.7cm]{Fig8a.eps} \hfill 
 \includegraphics[height=5.8cm]{Fig8b.eps} 
 \caption{
 Influence of the potential height $U_0$  on the transport exponent $\alpha_{\rm eff}$ (a)
 and thermodynamic efficiency $R_{\rm th,3}$ (b) as functions of loading force $f_0$
 for different sets of parameters:  $S_1$, $S_7$, $S_8$ (large cargo), 
 and $S_2$, $S_9$, $S_{10}$ (smaller cargo). 
 }
  \label{Fig8}
\end{figure}

With the increase of $U_0$ to
$25$ $k_BT_r$ and further to $30$ $k_BT_r$ thermodynamic efficiency indeed essentially 
increases. For the smaller cargo, it reaches a typical experimental value of 
50\% and even higher already for $25$ $k_BT_r$ at $f_0\sim 6- 7.5$ pN with 
the stalling force $f_0^{\rm stall}\approx 9.1$ pN, see in Fig. \ref{Fig8}, b.
The stalling force is a bit larger than for real kinesins. However, maximal thermodynamical
efficiency of about 58\% is also larger, as should be for a stronger motor. 
Nevertheless, this simple model
yields indeed realistic efficiencies and stalling forces at the same time. Moreover, 
our model motor can operate even in the regime of strongly anomalous transport with 
$\alpha_{\rm eff}\approx 0.58$ at the thermodynamic efficiency as large
as 70\%, for $U_0=30$ $k_BT_r$, at $f_0^{\rm opt}\approx 8.5$ pN with 
stalling force $f_0^{\rm stall}\approx 10$ pN, see in Fig. \ref{Fig8}, b. This is a real surprise! 
For smaller cargo the maximal efficiency is even larger, about 83\% at 
$\alpha_{\rm eff}\approx 0.92$, although this transport regime is close to normal.
Anomalous subdiffusive transport regime with such a huge efficiency, over 70\%,
was difficult to expect \textit{a priory}. 

\begin{figure}[t]
\vspace{1cm}
  \centering
  \includegraphics[width=10cm]{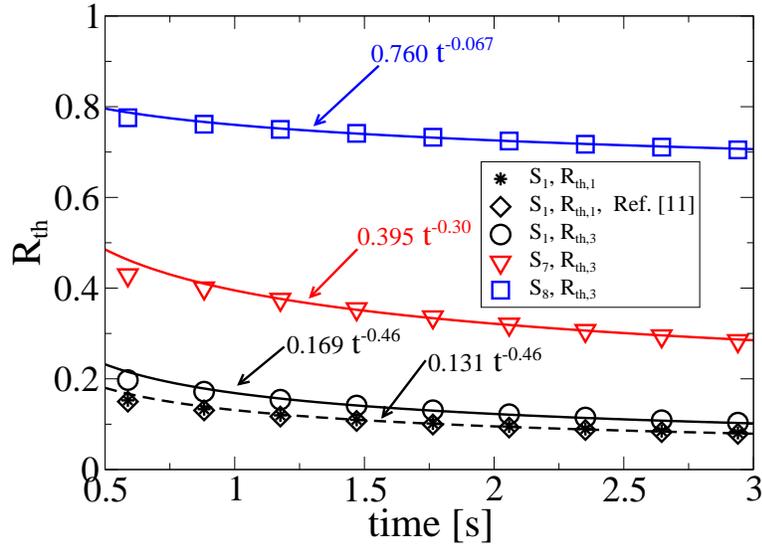} 
  \caption{Dependencies of the thermodynamical efficiencies on time and their power law fits
  for the sets $S_1$ 
(here and in Ref. \cite{PCCP14}), $S_7$, and $S_8$, with $f_0$ taken the values $3.07$,
 $5.12$ and $8.70$ pN, correspondingly. 
 }
  \label{Fig9}
\end{figure}

The explanation of this paradoxical
behavior reveals a profoundly new feature. Namely, the 
enzymatic cycling and the potential flashes occur in this case anomalously slow in time with
the power law exponent $\gamma$.  To understand this we plotted the time-dependence of
the thermodynamic efficiencies versus time in Fig. \ref{Fig9} for the sets $S_1$ 
(here and in Ref. \cite{PCCP14}), $S_7$, and $S_8$, at $f_0$ taken the values $3.07$,
$5.12$ pN, and $8.70$, correspondingly 
(near to the maximum of efficiency \textit{vs.} $f_0$). 
For $S_1$, both 
$R_{\rm th,1}(t)\propto 1/t^{1-\alpha_{\rm eff}}$, and 
$R_{\rm th,3}(t)\propto 1/t^{1-\alpha_{\rm eff}}$,
with $\alpha_{\rm eff}\approx 0.54$ confirming that $E_{\rm in, 1,3}(t)\propto t$.
However, for $S_7$ and $S_8$, $R_{\rm th,3}(t)\propto 1/t^{\lambda}$ with
$\lambda \neq 1-\alpha_{\rm eff}$. Assuming that $E_{\rm in, 3}(t)\propto t^\gamma$, one
obtains $\lambda=\gamma-\alpha_{\rm eff}$, from which $\gamma=\lambda+\alpha_{\rm eff}$.
Hence, from the data in Fig. \ref{Fig9} we deduce that $\gamma\approx 0.62$ 
($\alpha_{\rm eff}\approx 0.556$) for $S_8$
and $\gamma\approx 0.87$ ($\alpha_{\rm eff}\approx 0.57$) for $S_7$.
The occurrence of this thermodynamically highly efficient 
anomalous transport regime, where both the mean transport distance 
and the mean number of motor turnovers grow sublinearly in time, but with different
exponents, presents a profound result of this work, beyond recent
treatment in Refs. \cite{PLoSONE14,PCCP14}.

\section{Summary and Conclusions}

In this paper, we further generalized our model of anomalous transport in viscoelastic
cytosol of living cells realized by molecular motors like various kinesins. 
In this model, normally (in the absence of cargo) operating motor is pulling subdiffusive
(if not coupled to motor) cargo on an elastic linker, or tether. Subdiffusion
is described within non-Markovian GLE approach and its Markovian multi-dimensional 
embedding realization within a generalized Maxwell-Langevin model of viscoelasticity. 
The generalization
consisted in two aspects. First, we took the mechano-chemical coupling
between the motor cyclic turnovers and its translational motion
into account within a variant of the model of hand-over-hand motion of kinesin 
which was introduced in Refs. \cite{AstumianBier,Julicher2,Parmeggiani}. It is featured by
spatially-dependent rates of conformational transitions. This spatial dependence reflects 
biochemical
cycle kinetics of the molecular motor 
moving in a periodic binding potential. 
Our particular model choice was done in accordance with a biophysically plausible 
requirement that
ATP binding to the motor and its hydrolysis can be realized with the same rate $\alpha_1$ anywhere
on microtubule.  
This model choice allowed  comparison
with the ratchet model in Ref. \cite{PCCP14}, based on the same requirement,
by matching $\alpha_1$ 
with the doubled enzyme turnover rate in \cite{PCCP14}, using other parameters
the same and for $\Delta G_{\rm ATP}=U_0=20$ $k_BT_r$.
Second, we considered anharmonic linker with a maximally possible extension
length within the FENE model. This model choice allowed a direct comparison with
the purely elastic linker model in \cite{PCCP14} for the motor-cargo distances less than about one
fifth (with 4\% accuracy) of the maximal extension length. As a major result
of this study, we confirmed within the present more realistic setting, for realistic
model parameters, that all the major effects revealed in Refs. \cite{PLoSONE14,PCCP14} 
not only survives, but also quantitatively are very similar to the results in \cite{PCCP14}.
This allows to explain how the same motors operating in the same cells can realize
both normal and anomalous transport of various cargos depending on the cargo size,
strength of the motor (maximal or stall loading force 
which depends on the amplitude of binding potential), motor operating frequency
(which depends on the ATP binding  and hydrolysis rate), and the loading force opposing
the motion. 

However, an important discrepancy between two discussed models emerges on the
level of thermodynamic efficiency. Within the present model, the input energy fueling 
the motor operation is calculated as the energy required to accomplish  biochemical cycles
of the motor in the working direction by hydrolysing ATP molecules 
\cite{Julicher1,Julicher2,Parmeggiani}, in accordance with the main principles
of the free-energy transduction in isothermal engines \cite{Hill}, rather than
energy invested into the potential flashes unidirectionally \cite{Sekimoto}. The former is
less than the latter because of the energy recuperation (due to bidirectional coupling).
This makes thermodynamic efficiency of molecular motor essentially larger than one obtains in simple
ratchet models with unidirectional coupling and constant flashing rates. We showed that
our model can consistently explain near to normal transport with thermodynamic efficiency
of 50\% in viscoelastic environment of biological cells, for realistic parameters.
As a major surprise, we showed that a strongly anomalous subdiffusive transport is also
possible with thermodynamic efficiencies as high as 70\%. Here we revealed
a very important new feature. Namely, the biochemical enzyme turnovers  can become 
anomalously slow, $\langle N_{\rm turn}(t)\rangle 
\propto t^\gamma$, $0<\gamma<1$, due to mechanical coupling, not being characterized by a turnover rate
anymore. To the best of our knowledge, this is the
first time when anomalous enzyme kinetics of this kind, i.e. no mean turnover
rate exists, is obtain within
a physical approach based on fundamental principles of statistical mechanics \cite{Kubo,Zwanzig}.
To reveal such a regime provide a real challenge for experimental biophysicists.

It is important to mention that the difference in thermodynamic efficiencies
does not affect the major results in Refs. \cite{PLoSONE14,PCCP14} 
because normally such motors as kinesins are operating at zero thermodynamic efficiency
just relocating cargos from one place in the cell to another one, not increasing their potential
energy.  

Furthermore, we showed that the linker anharmonicity practically does not introduce any significant
difference in  the case of strong linkers with elastic constant typically used in biophysical 
literature \cite{Kojima}. However, a recent experiment \cite{Bruno} suggested that the elastic constant can be an order
of magnitude lower in viscoelastic environment of living cells as compare
with one in water. In Ref. \cite{PCCP14}, we
showed within the model of harmonic linker that the transport of large cargos is hardy possible
on such a weak linker when the motor operates at a high turnover frequency of about 100 Hz. 
The linker should then become broken.
However, in the present work we demonstrate that a weak linker can yet sustain such
a transport due to strong anharmonic effects. Moreover, we reaffirmed the emergence of a paradoxical regime of cargo's 
supertransport with $\alpha_{\rm eff}>1$ on a weak linker for the motor stepping normally with
$\alpha_{\rm eff}=1$ at its small operating frequencies.

To conclude, we hope that the further confirmation of the major results of  \cite{PLoSONE14,PCCP14} 
in a more realistic setup of this work, as well as new results of this work,
will inspire the followup experimental work,
which will
provide a further feedback to theoretical description of both anomalous and normal 
transport processes in the viscoelastic crowded environment provided by the cytosol of
living cells.

\ack
Support of this research by the German Research Foundation, Grant GO 2052/1-2, 
is gratefully acknowledged. \\

\end{document}